\documentstyle[citation]{article}
\font\tenbf=cmbx10
\font\tenrm=cmr10
\font\tenit=cmti10
\font\elevenbf=cmbx10 scaled\magstep 1
\font\elevenrm=cmr10 scaled\magstep 1
\font\elevenit=cmti10 scaled\magstep 1

\font\ninerm=cmr9

\def\epm{e^\pm}
\def\ep{e^+}
\def\em{e^-}
\def\wpm{W^\pm}
\def\wmp{W^\mp}
\def\wp{W^+}
\def\wm{W^-}
\def\wl{W_L^{}}
\def\wt{W_T^{}}
\def\pt{p_T^{}}
\def\mww{M^{}_{WW}}
\def\mw2{M_W^2}
\def\mh2{m_H^2}
\def\fbi{{\rm fb}^{-1}}
\def\to{\rightarrow}
\def\L{{\cal L}}
\def\M{{\cal M}}

\def\ss{{\sqrt s}}
\def\cos{{\rm cos}}
\def\Tr{\mathop{\elevenrm Tr}\nolimits}

\let\int=\intop         
\def\gsim{\buildrel {\mbox{$>$}} \over {\raisebox{-0.8ex}{\hspace{-0.05in}
$\sim$}}}
\def\lsim{\buildrel {\mbox{$<$}} \over {\raisebox{-0.8ex}{\hspace{-0.05in}
$\sim$}}}
\def\overlay#1#2{\ifmmode%
\setbox0=\hbox{$#1$}%
\setbox1=\hbox to\wd0{\hss$#2$\hss}\else%
\setbox0=\hbox{#1}%
\setbox1=\hbox to\wd0{\hss#2\hss}\fi%
 #1\hskip-\wd0\box1 }
\def\ptm{\overlay{/}{p}_T^{}}

\renewenvironment{thebibliography}[1]
 { \elevenrm
   \begin{list}{\arabic{enumi}.}
    {\usecounter{enumi} \setlength{\parsep}{0pt}
     \setlength{\itemsep}{3pt} \settowidth{\labelwidth}{#1.}
     \sloppy
    }}{\end{list}}

\begin{document}
\hfill\vbox{\hbox{FERMILAB--CONF--93/217--T}\hbox{July 1993}}\par
\begin{center}
\vglue 0.2cm

 {{\elevenbf        \vglue 10pt
              STRONGLY-INTERACTING ELECTROWEAK SECTOR\\
                 \vglue 3pt
                 AT FUTURE COLLIDERS
\footnote{\ninerm
Invited talk at the ``2$^{nd}$ International Workshop on

Physics and Experiments at Linear $e^+e^-$ Colliders",

Waikoloa, Hawaii, April 26-30, 1993.}
\\}
\vglue 5pt
\vglue 0.5cm
{\tenbf Tao Han \\}
\baselineskip=13pt
{\tenit Fermi National Accelerator Laboratory, P. O. Box 500\\}
\baselineskip=12pt
{\tenit Batavia, IL 60510, USA\\}}

%
\vglue 0.5cm
{\tenrm ABSTRACT}
 \end{center}
 \vglue 0.2cm
{\rightskip=3pc
 \leftskip=3pc
 \tenrm\baselineskip=12pt
 \noindent
If there are no light Higgs bosons found below $\cal O$(800 GeV) or so,
the interactions among longitudinally-polarized vector bosons ($\wl$) will

become strong at the TeV region, and new physics that is responsible for

the electroweak symmetry breaking must emerge at this energy scale. We
discuss the phenomenological prospects of the Strongly-interacting

Electro-Weak
Sector (SEWS) at future TeV linear colliders and hadronic supercolliders.
\vglue 0.5cm}
%
\baselineskip=14pt
\elevenrm

{\elevenbf\noindent 1. $W$-Boson Physics: from ``Weak'' to ``Strong''}
\vglue 0.5cm

This year, 1993, is the twentieth anniversary of the experimental
confirmation of the weak neutral current. It also marks the tenth
anniversary of the observation of the $W$ bosons (here and henceforth
$W$ generically denotes the $W^\pm$ and $Z$ bosons,

unless specified otherwise). These

discoveries demonstrated the major triumph of the $SU(2)_L\otimes U(1)_Y$

electroweak gauge structure. High energy experiments in the past two

decades have further verified the validity
of the Standard Model (SM) to high precision, including a constraint
on the top-quark mass $m_t = 150^{+19+15}_{-24-20}$ GeV\cite{Lang}.

We thus have been enjoying a beautiful theory that successfully describes

all particle phenomena up to highest energy scale accessible today.

However, among several unanswered fundamental questions in the Standard
Model, the elusive neutral scalar, the Higgs boson ($H$),

predicted by the SM as the remnant of the spontaneous

electroweak symmetry breaking, has not been observed.

In fact, there is no direct experimental evidence so far in favor of

any specific proposal for the mechanism of the electroweak symmetry

breaking. This is clearly one of the most prominent mysteries of
contemporary particle physics. Exploring the electroweak symmetry
breaking sector is one of the most challenging tasks for high energy
physicists, and it is one of the major motivations to build

the next generation of colliders.

To study the electroweak symmetry breaking sector,

it is instructive to first recollect the physics of weak bosons.
One of the earlier examples of conventional weak interactions

is the neutrino-nucleon scattering, $\bar \nu_e^{} + p \to n + e^+$.
It is described well by the four-fermion
contact interaction, and the amplitude goes like $G_F^{}s$, where
$G_F^{}$ is the Fermi weak-coupling constant and of order $10^{-5}$
GeV$^{-2}$, and $s$ the squared c.m. energy.

This amplitude would not respect unitarity at high energies.
\begin{figure}[t]
\vspace{1.6in}
\caption{\label{weak}
\tenrm \baselineskip=12pt
Representative Feynman diagrams for (a) low energy virtual $W$ effect
in   $\bar \nu_e^{} + p \to n + e^+$,
and (b) nearly on-shell $W^+$ production at high energies.}
\end{figure}
In the SM language,

it is a neutrino-up-quark scattering via a highly off-shell

$W$, as shown in Fig. 1(a). If $s \ll M_W^2$,
the effective coupling reduces to the conventional one,

$G_F^{}/{\sqrt 2} = g^2/8\mw2$,
where $g$ is the $SU(2)_L$ coupling. We see that the smallness of the

$G_F^{}$ value is nothing but the large-$M_W$ suppression at the low

energy. At collider energies, a nearly on-shell
$\wp$ production and decay can be described by the

same amplitude via $\bar d u$ annihilation (see Fig. 1(b))
as was already experimentally observed at the CERN


$ S p \bar p S$

and the Fermilab Tevatron.
The effective coupling is characterized by $g$, and the $W$-propagator
presents a proper high energy behavior.

We have now learned two things: first, the
conventional weak interactions are not intrinsically ``weak'': they are weak
because the energy available is much smaller than $M_W$; second,
the unitarity violation of four-fermion interactions at high energies
is cured by introducing the intermediate vector bosons. The scale of

$M^{}_W$ is the {\elevenit first} threshold of the ``weak'' interactions.

If we keep going to an even higher energy scale at next generation of
supercolliders, at which the squared c.m. energy $s \gg \mw2$,

the polarization vector of a longitudinally-polarized

vector boson ($\wl$) can be approximated as
$\epsilon_L^\mu \simeq p^\mu/M_W$, where $p$ is its momentum, and the
amplitude of pure $\wl \wl$ scattering goes like $s/v^2$, where

$v \simeq 246$ GeV,
is the vacuum expectation value of the scalar field. The interactions

among the longitudinal weak bosons would therefore become

strong and violate unitarity at about 1 -- 2 TeV\cite{lqt,changail}.
Something must happen before or at this energy scale to cure the bad high

energy behavior of $\wl\wl$ scattering and this will be the {\elevenit second}

threshold of the ``weak'' interactions\cite{lqt}.

It is the Higgs boson in the SM that comes in
to rescue the situation: Higgs-boson exchange diagrams cancel the

linear dependence on $s$, replaced by $\mh2$.
This can be easily understood by invoking the electroweak

``Equivalence Theorem''\cite{lqt,changail,cornwall},

which states that at high energies,
the external longitudinal vector bosons in the scattering amplitudes

recall their origin and
can be replaced by the corresponding Goldstone bosons ($w$'s).

Since the self-coupling $\lambda$ among the scalars (the Goldstone and

the Higgs bosons) in the SM is proportional

to $m_H^2$, the scattering amplitude of $\wl\wl$,

equivalent to that of the Goldstone bosons as symbolically shown

in Fig. 2, goes like $\mh2/v^2$. In the SM, $m_H^{}$ is essentially

a free parameter and the current experimental lower limit on

$m_H^{}$ from LEP I data is about 60 GeV\cite{mhiggs}.

Therefore, if there is no Higgs
boson found below 800 GeV or so, the $\wl\wl$ scattering nevertheless

becomes strong\cite{lqt,changail,dicusm,strng}

regardless what kind of new physics appears.
This is the scenario of the Strongly-interacting Electro-Weak
Sector (SEWS)\cite{reviews,baggeretal} that we will consider in this talk.

\begin{figure}[t]
\vspace{1.8in}
\caption{\label{scattering}
\tenrm \baselineskip=12pt
Symbolic diagrams for (a) longitudinally-polarized $\wl\wl$ scattering,

and (b) equivalent Goldstone-boson $ww$ scattering.}
\end{figure}

How do we expect to experimentally search for the SEWS? Obviously,
any information about the SEWS must be from studies of the massive
vector-bosons ($W$'s). Since we have been able to produce a large

number of $W$'s, especially a million $Z$'s at LEP I, the precision
measurements of electroweak parameters may provide important
information about the higher-scale physics beyond the

SM\cite{stu,morestu,tclimits}.

Indeed, from the ``oblique'' corrections for certain types of Technicolor

models, significant limits can be found.

However, these are not sensitive enough to
other cases, such as those with scalar Higgs bosons\cite{dkennedy}.

One can also consider to study the low energy effects from SEWS
on vector-boson self-interactions, such as the triple-vector-boson

vertices\cite{vvv,xlbdv,talkme}. However, as argued in Einhorn's

talk\cite{talkme}, the expected deviation from the SM values of the

vector-boson self-interactions may be only at a percentage level or less
since the dependence on the cutoff where the new physics enters is
at most logarithmic. Of course,

the most direct way of studying the SEWS is to look at the

$\wl\wl \to \wl\wl$ scattering at high energies,

because these processes couple to the SEWS most strongly.

In what follows, we will  mainly discuss the phenomenology of
$\wl\wl$ scattering at future colliders.

Before we proceed, two remarks are in order. First,

an electroweak sector with light Higgs bosons is a weakly-coupled

theory. The SM with a light Higgs boson is the simplest example
of this type.  The minimal supersymmetric model is another.

These are clearly very attractive options and

have been studied intensively both theoretically and phenomenologically,
although there are still theoretical difficulties in these models.
For recent reviews, see talks by Gunion and Kane in these

proceedings\cite{lighth}. However,

before a light Higgs boson is found, or any other strong experimental
evidence is established, such as in searches of SUSY particles,

we cannot simply ignore this seemingly more complex scenario of the
SEWS, especially when guiding our experimental designs.

After all, low-energy hadron physics is messy and QCD in that energy

regime is not as simple as one could have hoped. But Nature has been
mean to us at least once!

Secondly, it has been argued that the SM is not a

consistent effective theory if $m_H^{} \gsim $800 GeV or so\cite{trivial},
based on the triviality argument and lattice simulations.

We may nevertheless discuss the SM and its generalization
with a TeV Higgs boson as a prototype for models with SEWS.

In the following,

we first review in Section 2 some representative models and schemes in which

the SEWS occurs. We then discuss the feasibility of experimentally studying
the SEWS at future linear colliders and hadronic supercolliders in

Sections 3 and 4, respectively. We summarize in Section 5.

\vglue 0.5cm
{\elevenbf\noindent 2. Models and Schemes in Studying SEWS}
\vglue 0.5cm
There is little experimental guidance thus far on the underlying dynamics

responsible for the generation of mass. However, there is
a piece of information that provides important understanding
to the EW sector. Experimentally, the $\rho$-parameter\cite{veltman}
is very close to unity, which implies that $M_W^{} \simeq M^{}_Z
\cos\theta_W^{}$
to rather high precision. This relation holds automatically if there
are only Higgs doublets in the model. In more general models, it is
naturally protected if there is a global $SU(2)$ symmetry.
In this talk, we will assume the so-called ``custodial'' symmetry\cite{siki}
of $SU(2)_V$ broken from global $SU(2)_L \otimes SU(2)_R$,

and that the corresponding Goldstone bosons, $w^\pm$
and $z$, form an ``isospin''-triplet, formally analogous to
$\pi^\pm$ and $\pi^0$ in the low-energy strong interactions.

In this section,  we first discuss some implications of SEWS.

We then discuss several representative models and schemes
which we will adopt in the later phenomenological analyses.
We will restrict our attention to models with spin $J=0$ and isospin

$I=0$ resonances (the Higgs-boson-like, called scalar-dominance),

and $(J,I=1,1)$ resonances (the techni-rho-like, called vector-dominance),

as well as nonresonant models in $\wl\wl$ scatterings.
\vglue 0.4cm
{\elevenit \noindent 2.1. Implications of SEWS}
\vglue 0.4cm

We believe that the Goldstone bosons transform under the $SU(2)_V$ as

a triplet. But we do not have any satisfactory theories that
are consistent with experimental constraints to describe the dynamics

of a Strongly-interacting Electro-Weak Sector, or the possibly strong
interactions among the Goldstone bosons. Furthermore, technically,

the perturbation theory may have broken down in the SEWS,
so that there is no reliable way of performing theoretical calculations.
Unitarization of $\wl\wl$ scattering amplitudes based on certain models
may be necessary, which makes the predictions unitarization-scheme dependent.
Experimentally, the conventional bump-hunting method may not work

in studying $\wl\wl$ scattering due to very broad resonances or even

structureless (nonresonance) in the invariant mass spectrum,

which are common in SEWS.

While we know very little about the SEWS indeed,
we do have some general information about scattering matrix

elements of the Goldstone bosons in the SEWS. First, based on group theory

arguments of $SU(2)_V$, all scattering matrix elements are determined
by only one amplitude function $A(s,t,u)$, where $s, t,$ and $u$ are
the kinematical variables. They are given by
\begin{eqnarray}
\label{crossings}
\M(W^+_LW^-_L \to Z_LZ_L)\ &=& \ \M(Z_LZ_L\to W^+_LW^-_L)\ =\ A(s,t,u)

\nonumber \\
\M(W^+_LW^-_L\to W^+_LW^-_L)\ &=&\ A(s,t,u)\ +\ A(t,s,u)

\nonumber \\
\M(Z_LZ_L \to Z_LZ_L)\ &=&\ A(s,t,u)\ +\ A(t,s,u)\ +\ A(u,t,s) \\
\M(W^\pm_L Z_L\to W^\pm_L Z_L)\ &=&\ A(t,s,u)
\nonumber \\
\M(W^\pm_LW^\pm_L \to W^\pm_LW^\pm_L)\ &=&\ A(t,s,u)\ + \ A(u,t,s)\ ,
\nonumber
\end{eqnarray}
which satisfy the crossing relations.
Secondly, at low energies when the c.m. energy is much smaller than the

minimum of the lowest resonance and the electroweak scale $4 \pi v$,

the scattering amplitudes obey

the low-energy theorems (LET)\cite{weincgh},
which are solely determined by the global symmetries before and after
the spontaneous symmetry breaking. With our assumption of $SU(2)_V$,

the LET amplitude is
\begin{equation}
A(s,t,u)\ = \ s/v^2\;.
\end{equation}



Finally, the amplitudes must respect partial wave unitarity

at all energies. That is about all we can safely say and

we are ignorant of the detailed physics.

An important fact is that the background processes to the SEWS effects
in the SM are in principle perturbatively calculable. So the search

strategy is to understand the SM expectations as well as we can;

to search for deviations from the SM prediction;
and then to extract the underlying dynamics.

\vglue 0.4cm
{\elevenit \noindent 2.2. The $O(2N)$ Model}
\vglue 0.4cm

This first model represents an attempt to describe a
Higgs boson in the nonperturbative domain

at $\cal O$(1 TeV).

In the SM, the scalar self-coupling $\lambda \sim m_H^2$.

A heavy Higgs particle corresponds to a large value
of $\lambda$.  For $m_H^{} \gsim$ 1 TeV, naive perturbation theory
breaks down, and one must take a more sophisticated approach.

One possibility for exploring the nonperturbative regime is to
exploit the isomorphism between $SU(2)_L \otimes SU(2)_R$ and
$O(4)$\cite{einhorn}.
Using a large-$N$ approximation, one can solve the $O(2N)$
model for an arbitrary value of $\lambda$, to leading order in $1/N$.
The theory can be characterized by a scale $\Lambda$ of the
Landau pole, below which it is a self-consistent effective theory.

Large values of $\Lambda$ correspond to small couplings
$\lambda$ and relatively light Higgs particles.  In contrast, small
values of $\Lambda$ correspond to large $\lambda$ and describe the
nonperturbative regime.

It is not hard to show that the $\wl \wl$ scattering amplitudes
respect the unitarity condition for all energies below $\Lambda$.
In the following phenomenological discussions, we will take $N=2$

and $\Lambda = 3$ TeV. If we parameterize the position of the pole by its
``mass'' $M$ and ``width'' $\Gamma$ through the relation
$s=(M - i \Gamma/2)^2$, then $M \sim 0.8$ TeV and

$\Gamma \sim 600$ GeV\cite{baggeretal}.

\vglue 0.4cm
{\elevenit \noindent 2.3. The Chirally-Coupled Scalar Model}
\vglue 0.4cm

The second model describes the low-energy regime of a technicolor-like
model whose lowest resonance is a techni-sigma, at about $\cal O$(1 TeV).

The effective Lagrangian
for such a resonance can be constructed using the techniques of Callan,
Coleman, Wess, and Zumino\cite{ccwz}.
The resulting Lagrangian is consistent
with the chiral symmetry $SU(2)_L \otimes SU(2)_R$, spontaneously broken
to the diagonal $SU(2)_V$. We will skip the details\cite{baggeretal}
of constructing the chiral Lagrangian, which contains interactions
among a new scalar field $S$ and the Goldstone fields.


%

%

%
%
%


%
%

There are two free parameters in the Lagrangian, which can be traded for
the mass ($M_S^{}$) and the width ($\Gamma_S$) of the $S$. If we assume
that $S$ dominantly decays to a pair of Goldstone bosons, the width is then
given by
\begin{equation}
\Gamma_S\ =\ {3 g^2_S M^3_S\over 32 \pi v^2}\ .
\end{equation}
For $g^{}_S = 1$, the $S$ reduces to the Higgs boson $H$ in the SM.

For $g^{}_S \ne 1$, however, the $S$ is {\elevenit not} a typical $H$.

It is simply an isoscalar resonance of arbitrary mass and width.

In either case, one must check that the scattering amplitudes

are unitary up to the energy of interest.

In what follows, we will choose $M_S =$ 1.0 TeV, $\Gamma_S =$ 350
GeV. These values give unitary scattering amplitudes
up to 2 TeV\cite{baggeretal}.

\vglue 0.4cm
{\elevenit \noindent 2.4. The Chirally-Coupled Vector Model}
\vglue 0.4cm

This example provides a relatively model-independent description of
the techni-rho resonance that arises in most technicolor theories.
As above, one can use the techniques of nonlinear realizations to
construct the most general coupling consistent with chiral
symmetry\cite{ccwz}. We will not present the detail

discussions\cite{baggeretal} here to construct the chiral Lagrangian
involving interactions among the new vector field
(the techni-rho) and Goldstone fields.

There are two free parameters in the Lagrangian,

$g^{}_V$ and $a$. They once again
can be traded for the mass ($M_V^{}$) and the width ($\Gamma_V$) of the
new vector field,
\begin{equation}
\label{massw}
M^2_V\ =\ a g_V^2 v^2\ , \ \ \Gamma_V \ =\ {a M_V^3 \over 192 \pi v^2} \ .
\end{equation}
Because of the chiral symmetry, these two parameters completely
define the theory.

In the limit $g^{}_V \to \infty$ and $M^{}_V$ finite,

it implies $a \to 0$. The techni-rho $V_\mu$ decouples from
ordinary fields, so that the SM is recovered.

The approach described above is essentially the same as the
so-called BESS model\cite{BESSo}. There are two minor differences

here. We have ignored a possible direct coupling term between the

techni-rho and the fermions, which  is called the parameter $b$. Also,

we parameterize the theory by the mass $M^{}_V$ and the width

$\Gamma_V^{}$, while they choose the mass and the new coupling
constant $g''(=2 g^{}_V)$. It is easy to work out the relationships
between these two different choices:
\begin{equation}
\Gamma^{}_V\; = \; {M^5_V \over {192\pi g^2_V v^4}}, \ \

g^2/g^2_V\; =\; 768\pi v^2 \mw2 \Gamma^{}_V/M^5_{V}\; .
\end{equation}

In what follows we will choose $M_V^{} =$ 1 (2) TeV,

$\Gamma^{}_V =$ 25 (700) GeV, called Vector 1.0 (2.0),
for studies at $\ep\em$ ($pp$) colliders.
These values preserve unitarity up to 3 TeV and
are consistent with experimental constraints\cite{besslimit,dominici}.

\vglue 0.4cm
{\elevenit \noindent 2.5. Chiral Lagrangian Approach -- Nonresonant}
\vglue 0.4cm

As seen in the discussions of the previous two examples,
effective field theories provide a useful formalism
to describe resonances in $W_LW_L$ scattering beyond the
Standard Model in a relatively model-independent way.
They also can be used to describe nonresonant models in which the $\wl\wl$
scattering occurs below the threshold for resonant production.
The effective Lagrangian description allows one to construct scattering
amplitudes that are consistent with crossing and chiral
symmetry\cite{norm,xlbdv}.

The most important effects at high energies can be found by considering
the Lagrangian for the Goldstone fields,
\begin{eqnarray}
\L_{\elevenrm Goldstone} \ & = & {v^2 \over 4}\,
\Tr \partial_\mu \Sigma^\dagger
\partial^\mu \Sigma \nonumber \\
 & & +\ L_1\,\bigg({v \over \Lambda}\bigg)^2\,
 \Tr(\partial_\mu\Sigma^\dagger \partial^\mu \Sigma)
 \ \Tr(\partial_\nu\Sigma^\dagger \partial^\nu \Sigma) \nonumber \\
& & +\ L_2\,\bigg({v \over \Lambda}\bigg)^2\,\Tr(\partial_\mu\Sigma^\dagger
\partial_\nu \Sigma)\ \Tr(\partial^\mu\Sigma^\dagger \partial^\nu \Sigma)\ ,
\end{eqnarray}
where $\Lambda \lsim 4 \pi v$ denotes the scale of the new physics.

The Lagrangian above describes new physics at energies below the
mass of lightest new particles. To order $p^2$ in the energy expansion,
only one operator contributes, and its coefficient is universal which
is determined by the low-energy theorems (LET).

All the effects of the new physics
are contained in the coefficients of the higher-dimensional operators
built from the Goldstone fields. To order $p^4$, there are two

additional operators that contribute to $\wl\wl$ scattering. In a sense,
this approach is a model-independent parameterization of the new physics.

The difficulty with this approach is that the scattering
amplitudes violate unitarity between 1 and 2 TeV.  This indicates
that new physics is near, but there is no guarantee that new resonances
lie within reaches of the next generation colliders.

The amplitudes have to be unitarized. For simplicity, we follow
Chanowitz and Gaillard \cite{changail}, take the LET amplitudes,

and unitarize them by saturating the unitarity limit when they reach the

bound $|a^I_\ell| < 1$ (called LET CG). This simple treatment is

numerically not much different from the $K$-matrix method\cite{baggeretal}.

%
\begin{figure}[t]
\vspace{4.1in}
\caption{\label{kaoru1}
\tenrm \baselineskip=12pt
Total cross sections for the $\ep\em \to f \bar f' WW'$ processes
in the SM with $m^{}_H=0$.}
\end{figure}

\vglue 0.4cm
{\elevenit \noindent 2.6. N/D Approach}
\vglue 0.4cm

Recently, Hikasa and Igi proposed a model-independent scheme
\cite{hikasaigi} to study SEWS. They construct amplitudes
for the $\wl\wl$ scattering with scalar or vector resonances
in a framework of self-consistent $N/D$ method. The amplitudes satisfy

unitarity,

analyticity, and approximate crossing symmetry. This scheme could provide
a convenient basis for phenomenological studies of SEWS below a
typical scale about $4\pi v \simeq 3$ TeV.

%
\begin{figure}[t]
\vspace{3.9in}
\caption{\label{kuriharawz}
\tenrm \baselineskip=12pt
Invariant mass distribution of $WZ$ in the $\ep\em \to e^\pm \nu_e W^\mp Z$

process for signals of $M_V^{}=$0.8, 1, and 1.2 TeV, and
the continuum SM background. }
\end{figure}

\vglue 0.5cm
{\elevenbf \noindent 3. SEWS  at TeV Linear Colliders}
\vglue 0.5cm

We now discuss the feasibility of studying the SEWS

at future TeV linear colliders.

\vglue 0.4cm
{\elevenit \noindent 3.1. Via $\wl \wl$~Fusion Processes}
\vglue 0.4cm

The $W^+_LW^-_L$~fusion process dominates the heavy Higgs boson
production at TeV $\ep\em$ colliders\cite{dawsonand,jfg}.

More recently, two
groups re-examined the heavy Higgs boson signal and backgrounds
with a full set of SM diagrams\cite{kuriharan,kaorukm}.

Figure~\ref{kaoru1} shows the expected total cross sections versus
$\sqrt s$ for the $\ep\em \to f \bar f' WW'$ processes\cite{kaorukm}
in the SM.  The results were obtained with $m_H^{}=0$, so that they
can be viewed as the irreducible SM backgrounds to the SEWS.

The final state $\wl$-pairs from the SEWS are central, and

populated in the large $\mww$ region. The transverse momentum of
the pair, $\pt(WW)$, is of order $M^{}_W$ (see Section 4 for

more discussion). To observe a heavy Higgs-boson signal from the

fusion processes,

the basic kinematical cuts are
\begin{equation}
\label{cutsee}
 M^{}_{WW} >500~{\rm GeV}, \ \pt(WW) >50~{\rm GeV},\  |\cos(\theta_W)|<0.6\; .
\end{equation}
Processes of $WW'$ plus an $e^\pm$ in the final state are

dominantly induced by nearly on-shell photons,
and the rates for these processes are larger.
To suppress these backgrounds, besides the cuts in Eq.~\ref{cutsee},
electron-vetoing has been found effective\cite{kaorukm}

and the requirement is that
\begin{itemize}
\item no energetic $\epm$ with $E_e>50$~GeV in central

region  $|\cos(\theta_e)|<|\cos(0.15~{\rm rad})|$.
\end{itemize}
With these cuts, they found that at $\ss=$ 1.5 TeV
the signals for $m_H^{}=$ 1 TeV
in the processes $\ep \em \to \nu_e \bar \nu_e \wp \wm$ and
$\ep \em \to \nu_e \bar \nu_e ZZ$ are

clearly above the SM continuum backgrounds.

For the case of $m_H^{} \to \infty$ (LET),

the signal is structureless in the $M_{WW}$

spectrum and the background rate is comparable.  Higher c.m.

energy or a few hundred $\fbi$ of integrated luminosity is needed
to see the statistical significance of the signal\cite{kuriharan,kaorukm,admh}.
\begin{table}[t]
\caption{\label{table1}
\tenrm \baselineskip=12pt
Number of $\wl\wl$ events calculated for linear colliders at
1.5 (2.0) TeV, assuming an integrated luminosity of 200 (300) $\fbi$. }
\baselineskip=14pt
\begin{center}
\begin{tabular}{|c||ccccc|}
\hline
 & $O(2N)$  & Scalar 1.0  &  Vec 1.0   &   LET  & Bckgrnds \\

\hline
$\ep\em \to \nu_e \bar \nu_e \wp_L \wm_L$

 & 320 (1050) &   166 (570)   &   72 (270)  &   48 (165) & 90 (255)  \\

\hline

$\ep\em \to \nu_e \bar \nu_e Z_L Z_L$

 & 220 (690) &    150 (480)  &   70 (204)  &   88 (330) & 74 (216)  \\
\hline
\hline

$\em\em \to \nu_e \nu_e \wm_L\wm_L$

 & 50 (153)  &  70 (225)  &   70 (204)     &   88 (330)  &  202 (525) \\
\hline
\end{tabular}
\end{center}
\end{table}


Kurihara reported their recent studies\cite{kurihara} on a scalar and
a vector resonance based on the self-consistent $N/D$ approach\cite{hikasaigi}.

%
%

%
After some kinematical cuts\cite{kurihara}

to separate the backgrounds,

he concluded that with $\sqrt s=$1.5 TeV and ${\cal L}=$200~$\fbi$, a
1 TeV scalar (via $\wp\wm$ final state) or a vector (via $W^\pm Z$

channel) can be observed as a resonance, with a few tens of events.
Figure~\ref{kuriharawz} shows
the invariant mass distribution for

a vector resonance in

$W^\pm Z$ channel at $\sqrt s =$1.5 TeV. For $M_V >$1 TeV, the background
in this channel becomes large and higher energies and luminosities would

be needed to see the SEWS effects. However, more selective kinematical

cuts may be helpful to improve the situation.
\begin{figure}[t]
\vspace{5in}
\caption{\label{wwzz}
\tenrm \baselineskip=12pt
Invariant mass distributions of $M_{VV}$ at 1.5 TeV for the processes
$\ep\em \to \nu_e \bar \nu_e \wp_L\wm_L$ (solid),

and $\nu_e \bar \nu_e Z_L Z_L$ (dashed), (a) $O(2N)$ model;

(b) chirally-coupled vector model; (c) LET amplitudes. }
\end{figure}

To have a more coherent picture, in Table~\ref{table1}
we present the expected number of events for $\wp_L\wm_L$ and $Z_LZ_L$

final states via $\wp_L\wm_L$ fusion for models discussed in Sec. 2.
%



%
The results are for $\ss=$ 1.5 (2.0) TeV, assuming that the corresponding
annual integrated luminosity is 200 (300) $\fbi$, which is

from a rescaling of a 500 GeV NLC with 20 $\fbi$ to keep
a roughly constant event rate for
$\sigma_{point}=4\pi\alpha^2 /3 s \simeq$100 fb$/(E_{cm}~TeV)^2$.
The kinematical cuts used here are listed in Eq.~\ref{cutsee}.
Since we have used the ``Effective $W$-boson

Approximation''\cite{ewas} for the signal calculation, the $\pt(WW)$
cut cannot be implemented directly. Instead, we take a 75\% efficiency
for this cut based on exact matrix element simulations.

The $Z_LZ_L$ fusion diagrams have been ignored due to the much smaller
neutral current coupling of $\epm$. The background estimates

are obtained from the results of $m_H^{}=0$

by Hagiwara {\elevenit et al.}\cite{kaorukm}.
Although the signal rates are not very large and

the backgrounds are comparable, it seems possible to observe

statistically significant signal for any of the models

at $\ss=$1.5 TeV with a few hundred $\fbi$ integrated luminosity.

Another interesting option is an $\em\em$ collider\cite{emem}.
The event rates for $\em\em \to \nu_e \nu_e \wm_L\wm_L$

for different models are also listed in Table~\ref{table1},
with the same kinematical cuts.

This is a ``pessimistic'' channel in the sense that there is
no resonant contribution to this channel in any model discussed.
For instance, in a scalar-dominance model, $\sigma(W^-_LW^-_L)$ is
significantly smaller than those of $W^+_LW^-_L$ and $Z_L Z_L$
final states. Note that for nonresonant channels,

$\sigma(W^-_LW^-_L \to W^-_LW^-_L)=\sigma(W^+_LW^-_L \to Z_L Z_L)$

due to the crossing symmetry of Eq.~\ref{crossings}.
The irreducible background (with $m_H^{}=0$) is estimated\cite{emems}
to be moderately larger than the signal, as given in Table~\ref{table1}.

If we assume that other backgrounds such as $\em\em \to \em \em \wp \wm$

(relevant for hadronic decay modes)  via nearly real photons can be effectively

suppressed by the $\pt(WW)$ cut plus an electron-veto\cite{kaorukm},

we see from the table that we may also be able to observe statistically

significant signal at $\ss=$1.5 TeV with a few hundred $\fbi$ integrated

luminosity.

We should point out that there is another potential background,

$\ep \em \to \epm \nu \wmp Z$, which is not included in Table~\ref{table1}.
The rate of this background is comparable to the signal with the cuts of
Eq.~\ref{cutsee} and the
electron-veto\cite{kaorukm}, and we hope to

separate it by distinguishing the $\wpm$'s from $Z$'s in the final state.
How well can we tell a $\wpm$ from a $Z$ experimentally at the TeV
$\ep\em$ colliders? Of course, the charged leptonic decay

modes of $e$ and $\mu$ can be easily used, but the

branching fractions are rather small. It has been suggested\cite{peskin}

to tag a $b$-quark from $Z \to b \bar b$, with a branching fraction
of 15\%; while there is essentially no
$b$ from $\wpm$-decay since the top-quark is heavier than the $\wpm$.

However, the largest branching fraction is the decay to light quarks

(about 70\%). It would be very advantageous if one can make use of
the hadronic decay modes. Assume that the
energy resolution of the hadronic calorimeter is\cite{burke}
$$\Delta E_{hadron}=35\% {\sqrt E} \oplus 2\% E\;,$$
then, naively, the uncertainty of the hadronic energy from the decay
of a 500 GeV $W$ is about 12 GeV.
This is as large as the intrinsic $W^\pm$-$Z$ mass difference, so that
it is difficult to differentiate $\wpm$'s from $Z$'s by measuring
the di-jet mass of the decay products. However,
if we can combine the tracking information of the hadrons at the same
time, great improvement in the di-jet mass construction should be possible.

If the SEWS effects are observed, it is then desirable to
identify different models to uncover the underlying dynamics.
We find once again that effectively distinguishing the $\wpm$'s from $Z$'s
is essential to serve this purpose.
This is demonstrated in Fig.~\ref{wwzz}, where
we plot the invariant mass distributions for the processes
$\ep\em \to \nu_e \bar \nu_e \wp_L\wm_L$ and

$\nu_e \bar \nu_e Z_L Z_L$ at 1.5 TeV for three models: (a) the $O(2N)$ model;
(b) the chirally-coupled vector model (Vector 1.0);

and (c) the LET amplitude. For a scalar-dominance model,

we expect that the $\wp_L\wm_L$ rate is larger than
the $Z_LZ_L$ mode, like the case of the SM Higgs boson,
although the resonance structure may not be clear.

There will be a significant resonant enhancement for a vector-dominance

model in the $\wp_L\wm_L$ mode, but not in that of $Z_L Z_L$, just like
$\rho^0 \to \pi^+ \pi^-$ only. However, if the resonances
are far from our reach, then the LET amplitudes go like

$-u/v^2$ for $w^+w^-$ and $s/v^2$ for $zz$, so that

$\sigma(w^+ w^- \to z z)/\sigma(w^+ w^- \to w^+ w^-)=3/2$.
The $Z_L Z_L$ rate is then larger than that of $\wp_L\wm_L$,
and more so in the central region. Measuring the relative yields of
$\wp_L \wm_L$ and $Z_L Z_L$ will reveal some hints for the SEWS.
Although it may be a challenge to separate hadronic $\wpm$ and $Z$ events,

it is clearly important in studying the SEWS.

It has been argued\cite{rscmg} that if there are many particles
other than the three Goldstone bosons ($w^\pm,z$) in the SEWS,

the {\elevenit elastic} scattering amplitudes for $\wl\wl \to \wl \wl$

may be significantly reduced. And the {\elevenit inelastic}

scatterings, although strong, may be lack of discernible resonant structure.
The large inelastic channels themselves may be difficult to detect at the
hadronic supercollider environment. This is the scenario of the ``Hidden

symmetry-breaking sector''.

However, in the rather clean environment at $\ep\em$ colliders, it is possible
to study the inelastic channels directly. Although very much model-dependent,
the final states in the inelastic
channels are most likely pseudo-Goldstone bosons ($\phi$'s),
which would dominantly decay to heavy fermion pairs that are
kinematically accessible.

$$
\wp_L\wm_L \to \phi_i \phi_j \to f_i \bar f_i, f_j \bar f_j.
$$
The experimental signature then will be
\begin{itemize}
\item large missing transverse momentum, $\ptm \simeq M_W$, resulting

from $\wp_L\wm_L$ fusion;
\item four high-$\pt$ jets with two pairs reconstructing the masses of

the parent pseudo-Goldstone bosons:

$M^{}_{f_i \bar f_i} \simeq M_{\phi_i}$,
$M^{}_{f_j \bar f_j} \simeq M_{\phi_j}$.
\end{itemize}

Therefore, if a good hadronic mass resolution can be achieved, in contrast

to the hadronic supercolliders, those events from inelastic channels in
the ``Hidden symmetry-breaking sector'' could be spectacular.

\begin{figure}[t]
\vspace{4.1in}
\caption{\label{dominicif}
\tenrm \baselineskip=12pt
90\% C. L. contours in the plane $(M_V,g/g'')$


for c.m. energies at 0.3,
0.5, 1 TeV, ${\cal L}=$20 $\fbi$ and $b=0$. The solid lines correspond
to the bound from the unpolarized $WW$ differential cross section,

the dashed lines to the bound from  all the polarized
differential cross sections $\wl\wl$, $\wt\wl$,
$\wt\wt$ combined with the $WW$ left-right asymmetries.
The lines give the upper bounds on $g/g''$.}
\end{figure}

\vglue 0.4cm
{\elevenit \noindent 3.2. Via $e^+e^-$ Annihilation Processes}
\vglue 0.4cm

$\wl\wl$ fusion processes can go through different resonant channels,

so that they provide direct studies on the underlying physics. However,
the major disadvantage of this type of processes is  the inefficiency

of using the machine energy due to the loss of energy carried out by

the spectator leptons after radiating $\wl$'s. It is conceivable to
make the full use of the collider energy for vector resonant channels.

If $\sqrt s \simeq M_V$, then the vector resonance $V$ can be produced
by $\ep\em$ annihilation via either the direct $e \bar e V$ coupling or

via $W-V$ mixing.

Dominici presented their updated study\cite{dominici} on the limits of
BESS model parameters at the NLC of $\sqrt s \sim$ 0.3--2 TeV.
The processes of $\ep\em \to \wp\wm$ and
$\ep\em \to f \bar f'$ have been studied and variables such as the
total cross sections, forward-backward and left-right asymmetries
are used to constrain the virtual $V$ effects.

Figure~\ref{dominicif} shows the 90\% C. L. upper limits on the

$(M_V,g/g'')$ parameter plane at different assumed collider energies,

where $g'' =  2 g^{}_V$ in Eq.~\ref{massw}. One sees that the NLC will be
fairly sensitive to the BESS model parameters.
\begin{figure}[t]
\vspace{3.7in}
\caption{\label{barklowf}
\tenrm \baselineskip=12pt
  95\% C. L. contour for $Re(F_T)$ and $Im(F_T)$ at
  c.m. energy 1.5 TeV and ${\cal L}=500~\fbi$.}
\end{figure}

Another possible way of exploring the SEWS is via the final state

interactions (FSI) of the $\wl$-pairs produced in $\ep\em$ annihilation.

The FSI can be described by an Omn\'es function\cite{otherfsi,peskin}

and the rescattering coefficient $F_T$ is given by
\begin{equation}
F_T = \exp [ {1\over \pi} \int^\infty_0 ds'

\delta(s')\{ {1\over{s'-s'-i\epsilon}} - {1\over s'}\}
\end{equation}
where $\delta(s)$ is the phase shift and characterizes the

dynamics\cite{peskin,otherfsi,barklow}.
Barklow updated his study on FSI via a techni-rho-like vector resonance
in the $\ep\em \to \wp\wm$ process\cite{barklow}. Figure~\ref{barklowf}
shows the SM expectation, a 95\% C. L. contour, and a techni-rho

contribution of different masses at $\sqrt s=$1.5 TeV and an integrated

luminosity of 500~$\fbi$. One should be able to clearly see the

contribution from a techni-rho with $M_V > \ss$, and even possibly to
distinguish the weakly-interacting SM from SEWS with a few years of data.

Basdevant {\elevenit et al.}\cite{basdvnt} recently argued that the above

analysis corresponds to a Vector-Meson-Dominance
coupling of a techni-rho resonance. Generically, the FSI will result

in destructive interference between diagrams with and without rescattering,
so that the differential cross section could develop a dip near the resonance

region, instead of a bump. This may make the experimental observation more

difficult and it deserves further study.
\begin{figure}[t]
\vspace{3.8in}
\caption{\label{jikiaf}
\tenrm \baselineskip=12pt
Total cross section of $\gamma \gamma \to ZZ$ versus $\gamma \gamma$

c.m. energy in units of fb for several heavy Higgs boson masses.}
\end{figure}

We have seen that the $\ep\em$ annihilation processes ({\elevenit e. g.}

the BESS model and the FSI) are more advantageous in searching for

the SEWS effects, because of the full use of the collider energy,
and correspondingly less backgrounds.

However, it is essentially limited to a vector-dominance model
of the $(I,J=1,1)$ channel.

In other words, we would not expect to see significant effects
in the annihilation channel if there is no neutral

vector resonance to contribute.

\vglue 0.4cm
{\elevenit \noindent 3.3. TeV $\gamma\gamma$ Colliders}
\vglue 0.4cm

\begin{figure}[t]
\vspace{3.8in}
\caption{\label{backgrnd}
\tenrm \baselineskip=12pt
Representative diagrams for backgrounds to the $\wl\wl$
signal: (a) electroweak processes; (b) lowest-order QCD processes,
with possible additional QCD-jet radiation; and (c) top-quark

backgrounds.}
\end{figure}

Due to the recently developed idea of back-scattered laser beam

technique\cite{ginzburgkst}, it is possible to build a

$\gamma\gamma$ collider with similar energy and luminosity to

the $\ep \em$ one. With appropriate choice of the $\gamma\gamma$

polarizations, one can probe the scalar-dominance model via
the $J=0$ channel,\cite{peskin,carlim}

to compensate the shortcoming in the $\ep\em$ annihilation.

The SEWS effects in $\gamma\gamma$ colliders mainly go through loop

contributions. Signal process $\gamma\gamma \to W^+_L W^-_L$

has been calculated\cite{boosj,chao}
and the tree level background $\gamma\gamma \to W^+_T W^-_T$

is found to be larger  by about three orders of magnitude.
So there is little hope to separate the $\wp_L \wm_L$ mode

from the total since there is no effective way of measuring the $\wpm$
polarization. One may consider to study the

$\gamma\gamma \to Z_L Z_L$ final state since
there is no tree-level background. Jikia\cite{jikia}

first carried out a nice calculation and found that $Z_T Z_T$ final state

is still larger than that of $Z_L Z_L$ from a heavy Higgs boson
by more than an order of magnitude,

as shown in Fig.~\ref{jikiaf},

due to the huge contribution from $W^\pm$-loop. One could
improve the situation by looking at the angular distribution of the
fermions from the $Z$ decays, since it goes like sin$^2\theta_f$ for
$Z_L$ and $1+$cos$^2\theta_f$ for $Z_T$ at the $Z$-rest frame
with respect to the $Z$ moving direction. But it is very difficult to
achieve an order of magnitude suppression for the $Z_T Z_T$ background.
This conclusion has been confirmed more recently\cite{mberger}.

\vglue 0.5cm
{\elevenbf \noindent 4. SEWS  At Hadronic Supercolliders}
\vglue 0.5cm

The next generation of hadronic supercolliders, such as the SSC (40 TeV)
and the LHC (16 TeV),

could provide effective c.m. energies as high as a few TeV. This is
ideal in searching for the SEWS.
Another advantage for hadronic supercolliders over $\ep\em$

colliders is that, due to the great variety of the partons (quarks and gluons)

inside the proton, there are many channels open simultaneously, with
different spin, weak-isospin and charge states. However, the major problem
with high-energy high-luminosity
hadronic colliders is the messy backgrounds. As a comparison,
we discuss the prospects of searching for the SEWS at hadronic supercolliders
in this section. More details can be found in

Ref.~9 as well as the talk presented by Cheung\cite{kingman}.

Due to the large QCD backgrounds,
we are forced to concentrate on the purely leptonic decay modes of the
final state $W$'s, namely the ``gold-plated'' events,
with $W^\pm \rightarrow \ell^\pm \nu_\ell$ and
$Z \rightarrow \ell^+\ell^-$ $(\ell = e,\mu)$.

The experimental signature is then
given by two or more isolated, charged leptons in the central rapidity
$(y(\ell))$ region, with large transverse momenta $(p_T^{})$.
Although clean, these gold-plated channels carry the price of
relatively small branching fractions for the purely leptonic $W$ decays.

The diagram for longitudinal vector boson scattering is given symbolically
in Fig.~\ref{scattering}. In the case of $pp$ collisions, the initial
$\wl$'s are radiated from light quarks inside a proton.

The major backgrounds are symbolically depicted in Fig.~\ref{backgrnd}.

\begin{table}[t]
\caption{\label{table2}
\tenrm \baselineskip=12pt
Leptonic branching fractions (BR), kinematical cuts,

and jet-tag, -veto efficiencies
of the signal, for $ZZ$, $W^+W^-$, $W^+W^+$, and $W^+Z$ channels

in the studies of  SEWS at the SSC (LHC).}
\baselineskip=14pt
\begin{center}
\begin{tabular}{|c|c|}
\hline
 $ZZ (BR=0.45\%) $           &           $W^+W^- (BR=4.7\%) $ \\
\hline
$|y_\ell| < 2.5$            &            $|y_\ell| <2$ \\
$p_{T}>40$ GeV          &            $p_{T\ell}>100$ GeV \\
$p_{TZ} > \frac{1}{4}\sqrt{M^2_{ZZ}-4M_Z^2}$ & $\Delta p_{T\ell\ell}>450$ GeV
\\
$M_{ZZ}>500$ GeV            &                 $\cos \phi_{\ell\ell} <-0.8$ \\
                            &                 $ M_{\ell\ell}>250$ GeV \\
tag: $E_j({\rm tag})>1(0.8)$ TeV  &      tag: $E_j({\rm tag})>1.5(1.0)$ TeV \\
\mbox{\hspace{0.3in}} $3<|\eta_j({\rm tag})| < 5$  &
\mbox{\hspace{0.3in}} $3<|\eta_j({\rm tag})| < 5$ \\
                    & veto: $p_{Tj}({\rm veto}) > 30$ GeV \\
                    & \mbox{\hspace{0.3in}} $|\eta_j({\rm veto})| < 3$ \\
tag eff.: $59(49)\%$          &   veto eff.: $57(40)\%$ \\
                           &   veto+tag eff.: $38(24) \%$ \\
\hline
 $W^+W^+ (BR=4.7\%)$           &           $W^+Z (BR=1.5\%)$\\
\hline
$|y_\ell| < 2$            &            $|y_\ell| <2.5$ \\
$p_{T\ell}>100$ GeV          &            $p_{T\ell}>40$ GeV \\
$\Delta p_{T\ell\ell}>200$ GeV &        $\overlay{/}{p}_T>75$ GeV \\
$\cos \phi_{\ell\ell} <-0.8$    &        $ p^{}_{TZ} > \frac{1}{4} M_T$ \\
$ M_{\ell\ell}>250$ GeV        &        $M_T > 500$ GeV \\
                              &      tag: $E_j({\rm tag})>2(1.5)$ TeV \\
                & \mbox{\hspace{0.3in}} $3<|\eta_j({\rm tag})| < 5$ \\
veto: $p_{Tj}({\rm veto}) > 60$ GeV & veto: $p_{Tj}({\rm veto}) > 60$ GeV \\
\mbox{\hspace{0.3in}} $|\eta_j({\rm veto})| < 3$ &
 \mbox{\hspace{0.3in}} $|\eta_j({\rm veto})| < 3$ \\
veto eff.: $69(58)\%$       &   veto eff.: $75(48)\%$ \\
                         & veto+tag eff.:  $40(20)\%$ \\
\hline
\end{tabular}
\end{center}
\end{table}

It is important to note that two spectator quarks always emerge in
association with the $\wl\wl$ scattering signal, but that spectators emerge
in only a subset of the irreducible backgrounds.
The spectator quarks usually appear in forward/backward regions, and have
an energy of order 1 TeV and a $p_T^{}$ of order $M_W/2$. It is therefore
possible to improve the signal/background ratio by tagging those
quark jets (in particular, continuum pair production processes do not
have a spectator quark jet at lowest order in perturbation
theory)\cite{jtone}.
While studies have shown that tagging two high $p_T^{}$ spectator jets
substantially enhances the signal/background ratio,
such double tagging proves to be too costly to the

signal\cite{jttwo,jtthree,wpwpb,hzz}.
It has been recently suggested that
tagging just one of these quarks as a single energetic jet
can be just as efficient in suppressing the backgrounds
that do not intrinsically require spectator jets,
and far more efficient in retaining the signal for a
heavy Higgs boson\cite{hzz,hww,hwz}.
Thus, to isolate the heavy Higgs and other types of strong $W_LW_L$
signals, we will apply such a forward jet-tag for most final state
channels\cite{hzz,hww,hwz,wpwpg,wpwpo}.

\begin{figure}[t]
\vspace{4.6in}
\caption{\label{sclrm}
\tenrm \baselineskip=12pt
Invariant mass distributions for
the $O(2N)$ model with $\Lambda=$3 TeV

for the ``gold-plated'' leptonic final states that arise
from the processes $pp \rightarrow ZZX$,
$pp \rightarrow W^+W^-X$, $pp \rightarrow W^+ZX$ and
$pp \rightarrow W^+W^+X$, at 40 TeV

and an annual SSC luminosity of 10~fb$^{-1}$.

The longitudinally-polarized signal is plotted above the
summed background. The mass variable of

$x$-axis is in units of GeV, and the bin size is 50 GeV.}
\end{figure}

\begin{figure}[t]
\vspace{4.6in}
\caption{\label{vctrm}
\tenrm \baselineskip=12pt
The same as Fig.~\ref{sclrm}, but for the
chirally-coupled vector with $M_V = 2$ TeV, $\Gamma_V = 700$ GeV.}
\end{figure}

\begin{figure}[t]
\vspace{4.6in}
\caption{\label{normf}
\tenrm \baselineskip=12pt
The same as Fig.~\ref{sclrm}, but for
the Nonresonant model unitarized following Chanowitz and Gaillard.}
\end{figure}

Furthermore, the initial $\wl$'s
participating in the $\wl\wl$ scattering have a $1/(p_T^2+M_W^2)^2$
distribution with respect to the quarks from which they are emitted.

This is to be contrasted, for instance, with $\wt\wt$ scattering where

the initiating $\wt$'s have a $p_T^2/(p_T^2+M_W^2)^2$ distribution with

respect to the emitting quarks. The softer $p_T$ distribution in the $\wl\wl$

case has two primary consequences.

First, the spectator quarks left behind tend to emerge with smaller

$p_T^{}$ and correspondingly larger
rapidity than those associated with the background processes
containing spectator jets and $\wl\wt$ or $\wt\wt$ pairs.
Therefore we will normally veto hard central jets to enhance the
signal/background ratio\cite{wpwpb,wpwpg,wpwpo}. Such a veto retains
most of the signal events. As a further bonus,
a central jet-veto is especially effective in suppressing
the reducible background from heavy quark production and decay.
The jets associated with this latter type of background populate a
more central region than do those from spectator quarks.
Secondly, the final $\wl\wl$ pair is likely to have
much more limited net transverse motion than $\wl\wt$ and $\wt\wt$
pairs produced through the various irreducible backgrounds.
We then expect the charged leptons from the decays of the two final
$\wl$'s to be very back-to-back in the transverse

plane\cite{wpwpg,wpwpo,msbmsc}.
This is due not only to the limited $p_T$ of the $\wl\wl$ system but also

to the fact that the bulk of the leptons emitted from each final $\wl$
will have a significant (and relatively similar) fraction of the $\wl$'s total
momentum. The latter fact also implies that the leptons will generally be
very energetic. A cut requiring that the leptons
appearing in the final state be very energetic and

very back-to-back will substantially
reduce all backgrounds, while being highly efficient in retaining
the $\wl\wl$ signal events.

In Table~\ref{table2}, we summarize the leptonic branching fractions,
selective kinematical cuts, and jet-tag, -veto efficiencies of

the signal, for the
$ZZ$, $W^+W^-$, $W^+W^+$, and $W^+Z$ channels in the studies of SEWS at
the SSC (LHC). The signal efficiencies are determined from the full

SM calculation with $m_H^{}=$ 1 TeV, and are used for other models,
assuming that the jet kinematics is essentially independent of

specific SEWS models.
\begin{table}[t]
\caption{\label{table3}
\tenrm \baselineskip=12pt
Number of events for the ``gold-plated'' leptonic final states
at the SSC (LHC) for an integrated luminosity of 10 (100) $\fbi$.}
\baselineskip=14pt
\begin{center}
\begin{tabular}{|@{\extracolsep{0.15in}}c||ccccc|}
\hline
        &    $O(2N)$   & Scalar 1.0  & Vector 2.0  &  LET CG   & Bckgrnds  \\
\hline
$ZZ$    &   5.2 (6.4)  &   6.2 (7.5) &   1.1 (1.4) & 2.6 (2.5) & 1.0 (1.0) \\
\hline
$\wp\wm$&   24 (19)    &    30 (26)  &  15 (8.0)   &  16 (9.2) &  21 (18)  \\
\hline
$\wp Z$ &    1.5 (1.1) &   1.8 (1.4) &  9.5 (4.8)  & 5.8 (3.2) & 2.5 (2.4) \\
\hline
$\wp\wp$&    7.1 (10)  &   8.2 (12)  &   7.8 (12)  &   25 (27) & 3.5 (6.2) \\
\hline
\end{tabular}
\end{center}
\end{table}
\begin{table}[t]
\caption{\label{table4}
\tenrm \baselineskip=12pt
Estimated number of SSC (LHC) years needed (if $<$ 10) to observe the

SEWS effects at a 95\% Confidence Level.}
\baselineskip=14pt
\begin{center}
\begin{tabular}{|@{\extracolsep{0.15in}}c||cccc|}
\hline
        &    $O(2N)$   & Scalar 1.0  & Vector 2.0  &  LET CG    \\
\hline
$ZZ$    &   3.0 (2.5)  &   2.2 (1.8) &   - (-)     & 4.0 (4.8)  \\
\hline
$\wp\wm$&   0.75 (1.0) &  0.5 (0.75) &  1.2 (3.8)  &  1.2 (3.0) \\
\hline
$\wp Z$ &   - (-)      &   - (-)     &  0.75 (1.8) & 1.8 (4.0)  \\
\hline
$\wp\wp$&    2.2 (2.2) &   2.2 (1.8) &   2.5 (2.0) &  0.25 (0.5) \\
\hline
\end{tabular}
\end{center}
\end{table}

It is informative to look at the invariant mass distributions of the $\wl\wl$

pairs in different channels for different models.

In Figures~\ref{sclrm}-\ref{normf}, we present the invariant mass
distributions and $WZ$ cluster transverse mass\cite{trmass}

($M_T^{}$) distribution for the $O(2N)$,

the chirally-coupled vector and a nonresonance

model respectively. One can clearly see the enhancements in $W^+W^-$
and $ZZ$ channels from the scalar-dominance model, in $WZ$ channel from the
vector-dominance model (Vector 2.0),

and in $W^+W^+$ channel from a nonresonant LET model.

The results for one-year run at the SSC (LHC), with the three types

of SEWS models, are shown in Table~\ref{table3}.
We see once again that different channels are sensitive
to different types of underlying physics.
The $W^+W^-$ and $ZZ$ channels have the largest
signal/background ratio for the scalar-dominance models, with
$\wp\wm$ having much larger rate. Vector-dominance models are
more likely to be discovered in the $WZ$ channel, and $W^+W^+$ are more
sensitive to nonresonant models.
To further quantify the observability of the signal over backgrounds,
Table~\ref{table4} shows the number of years needed to
observe a particular model for each  channel at a 95\% Confidence Level,
at the SSC (LHC) with an annual luminosity 10 (100) $\fbi$,
calculated by assuming Poisson statistics\cite{baggeretal}.

Based on the above discussion,

we conclude that within 1--3 years of running at the SSC (LHC),
it is possible to observe the SEWS effects via the $\wl\wl$ scattering in

the gold-plated decay modes. All $WW'$ channels must be studied
coherently in order to extract the underlying dynamics.

\vglue 0.5cm
{\elevenbf \noindent 5. Summary}
\vglue 0.5cm
If there are no light Higgs bosons found below $\cal O$(800 GeV) or so,
the interactions among longitudinally-polarized vector bosons will

become strong at the TeV region, and the new physics that is responsible for

the electroweak symmetry breaking must emerge at this energy scale. We have
discussed the phenomenological prospects of the SEWS at future linear colliders
and hadronic supercolliders.

At a 1.5 TeV $\ep\em$ collider with an annual integrated luminosity
of 200 $\fbi$, the most straightforward search is for a $(I,J=1,1)$

resonance $V$ produced from $\ep\em$ annihilation

via $W$-$V$ mixing or final state interactions.

With a few-year run, we should be able to scan over a substantial part
of the parameter space, including the LET case of $M_V \to \infty$.
Unfortunately, a TeV $\gamma\gamma$ collider may not be a good place
to probe the $(I,J=0,0)$ channel via $\wp_L\wm_L$ or $Z_LZ_L$ final states,

due to the huge $\wp_T \wm_T$ and $Z_T Z_T$ backgrounds.

$\wl\wl$ fusion processes provide direct access to a variety of channels.

At a 1.5 -- 2 TeV $\ep\em$ collider with 200 -- 300 $\fbi$ integrated
luminosity, it is very promising to
observe the SEWS effects via the
fusion processes. Due to the
inefficient use of the total energy, the signal rate is relatively
small and the backgrounds are comparable. More careful studies on
the backgrounds are needed before drawing conclusions.

We emphasize the importance of effectively distinguishing

the hadronic $W^\pm$ and $Z$ decays, in order to examine the underlying
dynamics by comparing the $\wp_L \wm_L$ and $Z_L Z_L$ processes, as well
as to separate certain backgrounds.

Due to the relatively clean environment

at $\ep\em$ colliders, the possibly large inelastic channels of $\wl\wl$

fusion  in the ``hidden symmetry-breaking sector'' could also be
studied via the hadronic final state. This would make the $\ep\em$

colliders complementary to hadronic supercolliders in exploring the SEWS.

At hadronic supercolliders such as the SSC (LHC), the effective c.m.

energies are higher, but the environment of the backgrounds is
messier. After sophisticated kinematical cuts and with a few

tens (hundreds) $\fbi$ integrated luminosity,

it is possible to observe the SEWS effects at the SSC (LHC). Due to
rather small signal rates, higher luminosity option would be
desirable if the faked backgrounds can be kept under control.

Searching for the SEWS seems to be a hard experiment.
However, in designing the next generation of colliders and detectors before

a light Higgs boson is found, one has to bear this logical possibility

of the SEWS in mind.

\vglue 0.5cm
{\elevenbf \noindent Acknowledgements \hfil}
\vglue 0.5cm
I would like to thank J. Bagger, V. Barger, K. Cheung,
J. Gunion, G. Ladinsky, J. Ohnemus, R. Phillips, R. Rosenfeld, C.-P. Yuan,
and D. Zeppenfeld for collaborations on the related subjects

in the past few years. I am also grateful to M. Chanowitz and K. Hikasa

for comments in preparing this talk. Helpful discussions with T. Barklow,

D. Burke, E. Boos, D. Dominici, K. Hagiwara, G. Jikia, K. Kanzaki,

Y. Kurihara, M. Peskin, C. Quigg,
G. Valencia, S. Willenbrock, E. Yehudai are also appreciated.

Finally, I would like to thank the conference organizers
for a stimulating and enjoyable meeting.

\vglue 0.5cm
{\elevenbf\noindent  References \hfil}
\vglue 0.5cm

\end{document}